# PDD CRAWLER: A FOCUSED WEB CRAWLER USING LINK AND CONTENT ANALYSIS FOR RELEVENCE PREDICTION.

Prashant Dahiwale[1], M M Raghuwanshi[2] and Latesh Malik[3]

[1]Research Scholar , Dept of CSE, GHRCE and Assist Prof Dept of CSE, RGCER,Nagpur, India
`prashantdd.india@gmail.com`
[2]Department of Computer Science Engineering, RGCER, Nagpur India
`m_raghuwanshi@rediffmail.com`
[3]Department of Computer science Engineering, GHRCE, Nagpur, India
`lateshmalik@gmail.com`

## ABSTRACT

*Majority of the computer or mobile phone enthusiasts make use of the web for searching activity. Web search engines are used for the searching; The results that the search engines get are provided to it by a software module known as the Web Crawler. The size of this web is increasing round-the-clock. The principal problem is to search this huge database for specific information. To state whether a web page is relevant to a search topic is a dilemma. This paper proposes a crawler called as "PDD crawler" which will follow both a link based as well as a content based approach. This crawler follows a completely new crawling strategy to compute the relevance of the page. It analyses the content of the page based on the information contained in various tags within the HTML source code and then computes the total weight of the page. The page with the highest weight, thus has the maximum content and highest relevance.*

## KEYWORDS

*Web Crawler, HTML, Tags, Searching Relevance, Metadata*

## 1. INTRODUCTION

The World Wide Web is a huge collection of data. The data keeps increasing week by week, day by day and hour by hour. It is very important to classify data as relevant or irrelevant in accordance with users query. Researchers are working on techniques which would help to download relevant web pages. Researchers say that the huge size of data results in low exposure of complete data while search is performed and it is predicted that only one third of the data gets indexed[1].The web is so large that even the number of relevant web pages that get download is too large to be explored by the user. This scenario generates the need of downloading the most relevant and superior pages first. Web search is currently generating more than 13% of the traffic to Web sites [2]. Many policies have been developed to schedule the downloading of relevant web pages for a particular search topic. Some of the policies are Breadth First Search, Depth First Search, Page Ranking Algorithms, Path ascending crawling Algorithm, Online Page Importance Calculation Algorithm, Crawler using Naïve Bayes Classifier, Focused Web Crawler, Semantic web Crawler etc. Each technique has its pros and cons. Focused Web Crawler is a technique which uses the similarity major to map relatedness among the downloaded page and unvisited page. This technique ensures that similar pages get downloaded and hence the name Focused web crawler[3]. Web crawler needs to search for information among web pages identified by URLs. If we consider each web page as a node, then the World Wide Web can be seen as a data structure that resembles a Graph'. To traverse a graph like data structure our crawler will need traversal mechanisms much similar to those needed for traversing a graph like Breadth First Search (BFS) or Depth First Search (DFS). Proposed Crawler follows a simple Breadth First Search'approach. The start URL given as input to the crawler can be seen as a start node' in the graph. The hyperlinks extracted from the web page associated with this link will serve as its child nodes and so on. Thus, a hierarchy is maintained in this structure. Each child can point to its parent is the web page associated with the child node URL contains a hyperlink which is similar to any of the parent node URLs. Thus, this is a graph and not a tree. Web crawling can be considered as putting items in a queue and picking a single item from it each time. When a web page is crawled, the extracted hyperlinks from that page are appended to the end of the queue and the hyperlink at the front of the queue is picked up to continue the crawling loop. Thus, a web crawler deals with the infinite crawling loop which is iterative in nature. Since crawler is a software module which deals with World Wide Web, a few constraints [4] have to be dealt with: High speed internet connectivity, Memory to be utilized by data structures, Processor for algorithm processing and parsing process Disk storage to handle temporary data.

## 2. RELATED WORK

Mejdl S. Safran, Abdullah Althagafi and Dunren Che in Improving Relevance Prediction for Focused Web Crawlers'[4] propose that a key issue in designing a focused Web crawler is how to determine whether an unvisited URL is relevant to the search topic. this paper proposes a new learning-based approach to improve relevance prediction in focused Web crawlers. For this study, Naïve Bayesian is used as the base prediction model, which however can be easily switched to a different prediction model. Experimental result shows that approach is valid and more efficient than related approaches.
S. Lawrence and C. L. Giles in Searching the World Wide Web'[1] state that the coverage and recency of the major World Wide Web search engines when analyzed, yield some surprising results. Analysis of the overlap between pairs of engines gives an estimated lower bound on the size of the indexable Web of 320 million pages. Pavalam S. M., S. V. Kasmir Raja, Jawahar M., and Felix K. Akorli in Web Crawler in Mobile Systems' [6] propose that with the advent of internet technology, data has exploded to a considerable amount. Large volumes of data can be explored easily through search engines, to extract valuable information. Carlos Castillo, Mauricio Marin, Andrea Rodriguez in Scheduling Algorithms for Web Crawling'[7] presents a comparative study of strategies for Web crawling. It shows that a combination of breadth-first ordering with the largest sites first is a practical alternative since it is fast, simple to implement, and able to retrieve the best ranked pages at a rate that is closer to the optimal than other alternatives. The study was performed on a large sample of the Chilean Web which was crawled by using simulators, so that all strategies were compared under the same conditions, and actual crawls to validate our conclusions. The study also explored the effects of large scale parallelism in the page retrieval task and multiple-page requests in a single connection for effective amortization of latency times.
Junghoo Cho and Hector Garcia-Molina in Effective Page Refresh Policies for Web Crawlers' [8] study how to maintain local copies of remote data sources fresh, when the source data is updated autonomously and independently. In particular, authors study the problem of web crawler that maintains local copies of remote web pages for web search engines. In this context, remote data sources (web sites) do not notify the copies (Web crawlers) of new changes, so there is a need to periodically poll the sources, it is very difficult to keep the copies completely up-to-date. This paper proposes various refresh policies and studies their effectiveness. First formalize the notion of Freshness of copied data by defining two freshness metrics, and then propose a Poisson process as a change model of data sources. Based on this framework, examine the effectiveness of the proposed refresh policies analytically and experimentally. Results show that a Poisson process is a good model to describe the changes of Web pages and results also show that proposed refresh policies improve the freshness of data very significantly. In certain cases, author got orders of magnitude improvement from existing policies. The Algorithm design Manual'[9] by Steven S. Skiena is a book intended as a manual on algorithm design, providing access to combinatorial algorithm technology for both students and computer professionals. It is divided into two parts: Techniques and Resources. The former is a general guide to techniques for the design and analysis of computer algorithms. The Resources section is intended for browsing and reference, and comprises the catalog of algorithmic resources, implementations, and an extensive bibliography. Artificial Intelligence illuminated' [10] by Ben Coppin introduces a number of methods that can be used to search, and it discusses discuss how effective they are in different situations. Depth-first search and breadth-first search are the best-known and widest-used search methods, and it is examined why this is and how they are implemented. A look is also given at a number of properties of search methods, including optimality and completeness[11][12], that can be used to determine how useful a search method will be for solving a particular problem.

## 3. DESIGN METHODOLOGY

In general, a web crawler must provide the features discussed [13] ,
1. A web crawler must be robust in nature Spider traps are a part of the hyperlink structure in the World Wide Web. There are servers that create spider traps, which mislead crawlers to infinitely travel a certain unnecessary part of the web. Our crawler must be made spider trap proof.
2. Web pages relate to web servers, i.e. different machines hosting these web pages and each web page has its own crawling policy, thus our crawler must respect the boundaries that each server draws.

### 3.1 Existing Methodology [14] [15] [16]

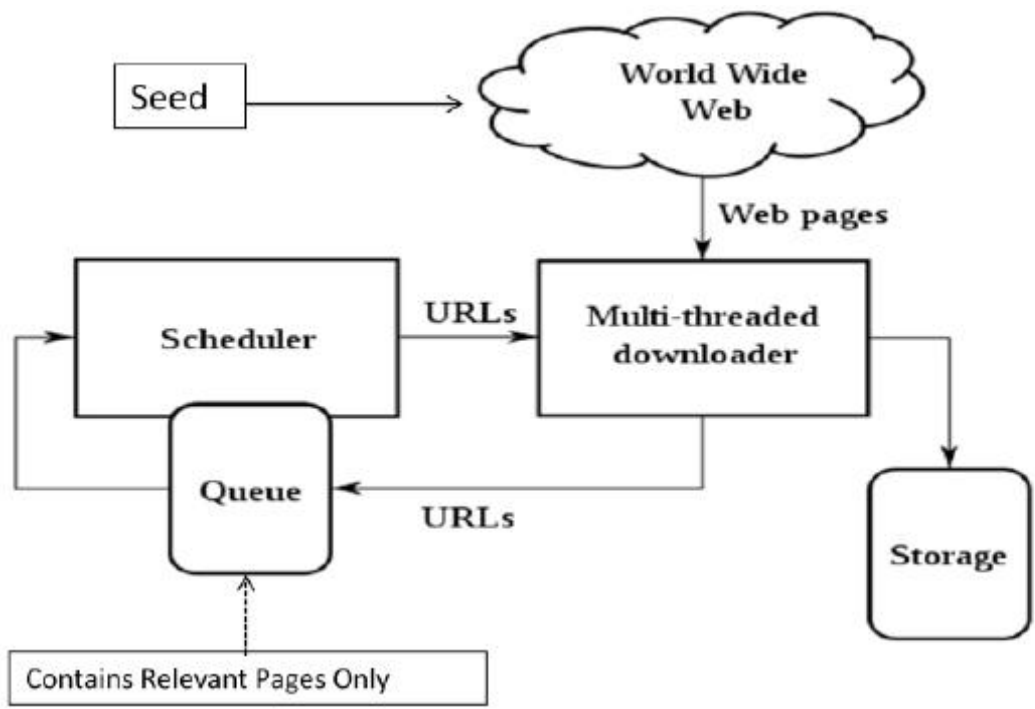

Figure1. High-level architecture of a standard Web crawler

Components of the web crawler are described as [4]:

1. Seed - It is the starting URL from where the where crawler starts traversing the World Wide Web recursively.

2. Multithreaded Downloader: It downloads page source code whose URL is specified by seed.

3. Queue: Contains unvisited hyperlinks extracted from the pages.

4. Scheduler: It is FCFS scheduler used to schedule pages in Queue.

5. Storage: Storage may be volatile or non-volatile data storage component.

### 3.2 Proposed Approach

1. This paper proposes a PDD crawler', which is both links based and content, based. The content that had been unused by the previous crawlers will also take part in the parsing activity. Thus, Rank Crawler is a link as well as content-based crawler.

2. Since true analysis of the web page is taking place (i.e. the entire web page is searched for the content fired by the user), this crawler is very well suited for business analysis purposes.

### 3.3 Architecture

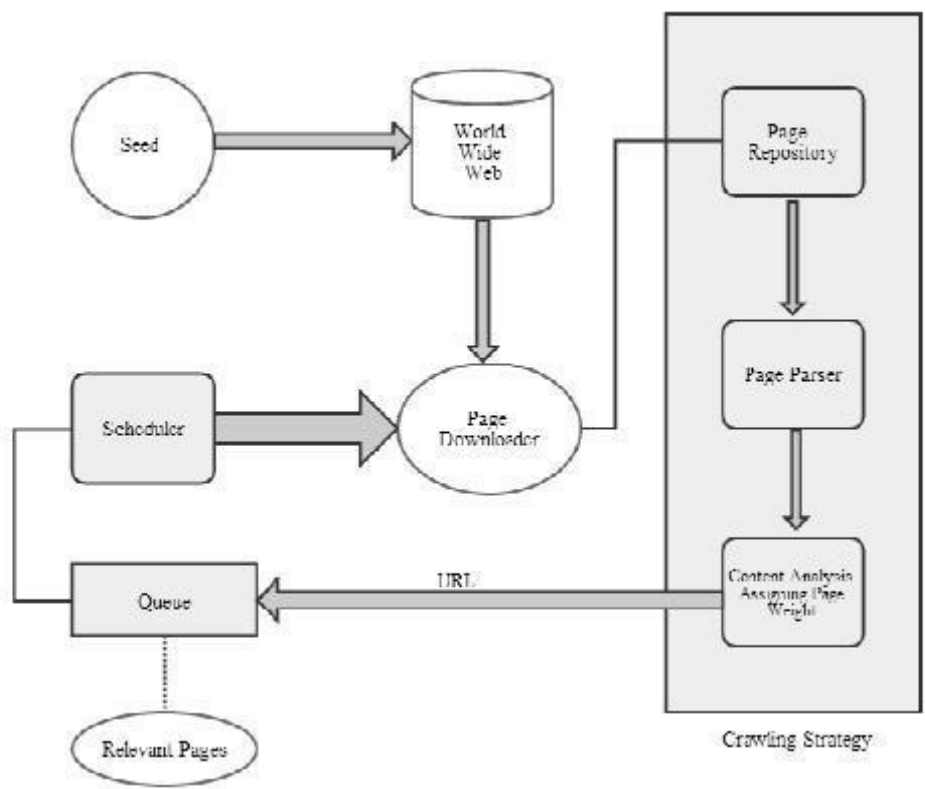

Figure 2. Proposed Architecture

Before describing the actual working of proposed Crawler, it is important to note that:
1. Every URL has got a web page associated with it. Web pages are in the form of HTML source code in which information is divided in various tags. Thus, page parsing and information retrieval are solely based on the tags of HTML source code.
2. Web pages have a certain defined structure, a hierarchy of sorts. This hierarchy defines the importance of tags with respect to a web page. Meta'is the most important tag or the tag with the highest priority, whereas body' is the one with the lowest priority.

Proposed PDD Crawler' follows these steps:

1. The seed URL used to connect to the World Wide Web. Along with the seed, a search string is to be provided by the user. The search string will act as the query that the user is searching for. If the URL is valid and the string is valid, then the crawler proceeds else it stops.

2. Every Web page has got tags that may contain information. Weights in the form of integers have been assigned to the tags of the page. If the content is found on the page (once or multiple times), the weight is to be multiplied by the number of occurrences in each tag and the total weight is added up to get the total page weight.

3. The Seed downloaded and all the URLs on the seed pages are extracted. The seed page is then checked to see whether it contains any relevant information. If yes, then the weight is calculated and it is set aside for further use, else the page is irrelevant to users query and has to be thrown away.

4. The URLs extracted are then scheduled to be downloaded one by one and the above process is followed (i.e. download a page and check its contents). If it returns a positive page weight, then set it aside and recursively do the same for every page.

5. The page with the most weight has the highest content. Thus, the results have to be descending sorted.

### 3.4 Architecture Components

1. Page Repository: Repository refers to the web pages that have been traversed and parsed. It may be a temporary storage from where the crawlers can refer the pages or it can be the browsers cache or storage from where it can refer the pages for faster access.

2. Page Parsing Mechanism: Page parsing mechanism takes pages one at a time from the page repository and searches for the keyword or the string in the page and based on that assign weight to the page.

3. Content Analysis: Final phase, which decides whether the page is relevant or whether it has to be discarded. Relevancy is decided on the basis of relevancy threshold which we have maintained in the algorithm.

### 3.5 Working of 'PDD Crawler

Proposed PDD Crawler' works using the source code of the page i.e. the downloader uses the source code to analyze the tags and contents of the page so as to get the page weight and calculate the degree of relevancy of a page. To make a simple web page you need to know the following tags:
1. < HTML > tells the browser your page is written in HTML format

2. < HEAD > this is a kind of preface of vital information that does not appear on the screen.
3. < TITLE > Write the title of the web page here - this is the information that viewers see on the upper bar of their screen.
4. < BODY > This is where you put the content of your page, the words and pictures that people read on the screen.
5. <META> This element used to provide structured metadata about a Web page. Multiple Meta elements with different attributes are often used on the same page. Meta elements can be used to specify page description, keywords and any other metadata not provided through the other head elements and attributes.

Consider the source code associated with the web page of the URL: **http://www.myblogindia.com/html/default.asp** as given below-

*<html>*
*<head>*
*<meta name="description" content="Free HTML Web tutorials">*
*<meta name="keywords" content="HTML, CSS, XML">*
*<meta name="author" content="RGCER">*
*<meta charset="UTF-8">*
*< title > HTML title of page< /title >*
*</head>*
*< body>*
*This is my very own HTML page. This page is just for reference.*
*< /body >*
*< /html >*

The downloader gets this content from a page. Downloading a page refers to the function of getting the source code of the page and analyzing as well as performing some actions on it. This method is what we call Parsing. The analysis part of the source code is as follows:

1. Let the Total weight of the page be t' units
2. The body tag has got the weight B' units
3. The title tag has got the weight T' units
4. The Meta tag has got the weight M' units
5. The heading tag (h1 through h6) has weight H' units
6. The URL has weight U' units
7. The no. of occurrence of the search string in body be Nb
8. The no. of occurrence of the search string in title be Nt
9. The no. of occurrence of the search string in META be Nm
10. The no. of occurrence of the search string in heading be Nh
11. The no. of occurrence of the search string in URL be Nu

The total weight of the page would be –

**t = (Nb*B)+(Nt*T)+(Nm*M)+(Nh*H)+(Nu*U)**

Assumptions for calculating page weight are defined below: M = 5 units, U = 4 units, T = 3 units, H = 2 units, B = 1 units Suppose the search string the user entered is: html (not case-sensitive). The number of occurrences of html in the following tags is as follows: Nb = 1, Nt = 1, Nm = 2, Nh = 0, Nu =1 The total weight of the page comes out to be:

$$t = (1*1) + (1*3) + (2*5) + (0*2) + (1*4)$$
$$t = 1 + 3 + 10 + 4$$
$$t = 18$$

Conclusion:
1. The page weight (t) > 3, thus the page is relevant.
2. Page is Relevant if and only if t > 3.
3. All pages with t<=3 will be discarded.

4. Will only work for static pages.
5. Static pages are the one which do not change or update or alter their content on regular basis i.e. the data change that is there on pages is either periodic or none what so ever.
6. Some pages are non-crawl able (e.g. the pages with META as robot).

## 4.RESULT

The proposed PDD Crawler' was implemented on the following hardware and software specifications: OS Name- Microsoft Windows 7 Ultimate, Version-6.1.7600 Build 7600, Processor- Intel(R) Core(TM) i5-3230M CPU @ 2.60 Ghz, 2601.and compared its performance with intelligent web crawler which is link based crawler , after running both the crawlers in same hardware and software environment with same input parameters, observed following rate of precisions for both the crawlers as shown in Table 1. evaluate the performance of proposed framework by comparing it with Intelligent web crawler by applying both the crawlers to the same problem domains such as Book Show, Book to read, Cricket match and Match making. The experiments are conducted on different social aspects database (web) having same words with different meaning as book meaning reserving some resource for the domain Book show‖ while as book meaning a physical entity made up by binding pages for the domain book to read. From the above test cases it is vibrant that if correct Seed URLs are provided according to domain sense of the word then:

1. The precision range comes out to be 20 to 70% which is fairly acceptable for the crawl.
2. The test results show that the search is now narrowed down and very specific results are obtained in sorted manner.
3. The sorting of the results cause the most relevant link to be displayed first to the user so as to save his valuable search time

Table 1: Result Comparisons

| Sr No | Seed URL | Query | Intelligent Crawler | PDD Crawler |
|---|---|---|---|---|
| | | | Percentage Precision | |
| 1. | www.bookmyevent.com | Book show | 22.14% | 59.37% |
| | www.ticketnew.com | | | |
| | www.bookmyshow.com | | | |
| 2. | www.bestsellers.about.com | Book to read | 33.64% | 67.53% |
| | www.publicbookshelf.com | | | |
| | www.goodreads.com | | | |
| 3. | www.espncricinfot.com | Cricket match | 32.43% | 65.87% |
| | www.starsports.com | | | |
| | www.icc-cricket.com | | | |
| 4. | www.astrosage.com | Match making | 42.65% | 69.89% |
| | www.mangalsutra.com | | | |
| | www.drikpanchang.com | | | |

4. The testing of the framework is carried out on live web, thus by precision reported accuracy of proposed Focused Web Crawler is promising.

Above test results completely depend on Seed URLs. The Seed URLs provide the correct domain for the search. Seed URLs help to initiate and guide the search in interested domain. Later on the Focused nature of the proposed Crawler will always deviate the search towards target links.

# 5 CONCLUSION AND FUTURE WORK

The main advantage of proposed crawler over the intelligent crawler and other Focused Crawlers is that it does not need any Relevance Feedback or training procedure in order to act intelligently; two kinds of change were found after comparing result of both the crawlers.
1. The number of extracted documents was reduced. Link analyzed, and deleted a great deal of irrelevant web page.
2. Crawling time is reduced. After a great deal of irrelevant web page is deleted, crawling lode is reduced.

In conclusion, after link analysis and page analysis in proposed crawler, crawling precision is increased and crawling rate is also increased. This will be an important tool to the search engines and thus will facilitate the newer versions of the search engines

**Authors**

**Prashant Dahiwale**: Is a Doctoral research fellow at G H Raisoni college of Engineering and research , Nagpur under RTM Nagpur University Nagpur in computer science and engineering. Having 5 years of teaching experience to teach M.tech and B.E comp Sc Engg courses and 1 year of industrial experience. His area of research is information retrieval and web mining. Completed Post PG diploma from CDAC in 2010. M.Tech in Comp Sc Engg form Visvesvaraya National Institute of Technology, (VNIT) Nagpur in 2009,B.E from RTM Nagpur university Nagpur in 2006. Published various research papers at International and National journals and conferences. Also author is a four time finalist of Asia regional ACM ICPC since 2010.organised various workshop on programming language fundamentals and advance concepts.

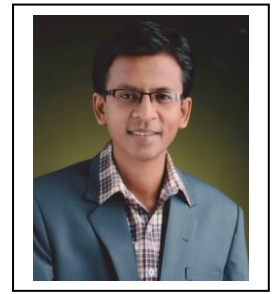

**Dr M M Raghuwanshi**: Professor dept of Comp sc Engg ,RGCER, Nagpur India . having total 25 plus years of teaching experience and 5 years of industrial experience. His research area is genetic algorithm, he has chaired multiple international and national conferences, he is IEEE reviewer, Ph D form Visvesvraya National Institute of Technology, Nagpur, and M.Tech from IIT Kharagpur. he is a author of text book on Algorithm and Data structure , He thought multiple subjects to M.Tech and B.E courses. he has organised multiple workshops on Theory of computations and Genetic algorithm. Published various research papers at International and National journals and conferences.

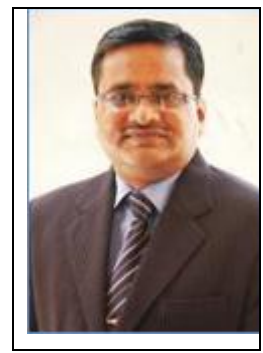